\documentclass[a4paper]{article}
\usepackage[utf8]{inputenc}
\usepackage{a4wide} %narrow margins
\usepackage{authblk}
\usepackage{url}
\usepackage[colorlinks=true,linkcolor=blue,citecolor=blue,urlcolor=blue]{hyperref}
\usepackage[pdftex]{graphicx,color}	%graphics
\usepackage{multirow}
\usepackage{soul}
\usepackage{amsmath}
  
\begin{document}

\title{Bovine Tuberculosis in Britain -- identifying signatures of polarisation and controversy on Twitter}
\author[1,*]{Christopher J.\ Banks}
\author[2]{Jessica Enright}
\author[3]{Sibylle Mohr}
\author[1,4]{Rowland R.\ Kao}
\affil[1]{\small Roslin Institute, University of Edinburgh}
\affil[2]{\small School of Computing Science, University of Glasgow}
\affil[3]{\small School of Biodiversity, One Health, and Veterinary Medicine, University of Glasgow}
\affil[4]{\small School of Physics and Astronomy, University of Edinburgh}
\affil[*]{\small Correspondence: \texttt{c.banks@ed.ac.uk}}

%\date{}
\maketitle

%%%%%%%%%%%%%%%%%%%%%%%%%%%%
\begin{abstract}
  Approaches to disease control are influenced by and reflected in public opinion, and the two are intrinsically entwined. Bovine tuberculosis (bTB) in British cattle and badgers is one example where there is a high degree of polarisation in opinion. Bovine viral diarrhoea (BVD), on the other hand, does not have the same controversy.

  In this paper we examine how language subjectivity on Twitter differs when comparing the discourses surrounding bTB and BVD, using a combination of network analysis and language and sentiment analysis. That data used for this study was collected from the Twitter public API over a two-year period. We investigated the network structure, language content, and user profiles of tweets featuring both diseases.

  While analysing network structure showed little difference between the two disease topics, elements of the structure allowed us to better investigate the language structure and profile of users. We found distinct differences between the language and sentiment used in tweets about each disease, and in the profile of the users who were doing the tweeting. We hope that this will guide further investigation and potential avenues for surveillance or the control of misinformation.
\end{abstract}

%%%%%%%%%%%%%%%%%%%%%%%%%%%
\section{Introduction}
Approaches to disease control are influenced by and reflected in public opinion, and the two are intrinsically entwined---disease control policy influences public opinion and public opinion influences disease control policy making. Polarisation of that opinion is therefore profoundly important and social media is well-known as a mechanism for enhancing polarisation.

While much of the literature on polarisation in social media focuses on political issues~\cite{conover2011,recuero2019,aldayel2022}, there is no reason why the principles cannot be extended to discourse on disease control. In this paper we will examine the indicators of controversy and polarisation in two agricultural diseases: bovine tuberculosis and bovine viral diarrhoea.

Bovine tuberculosis (bTB) in British cattle and badgers is one example where there is a high degree of polarisation in opinion. In the United Kingdom considerable heated debate centres around the legitimacy of badger culling as a control strategy for the disease in cattle~\cite{godfray2013}. Bovine viral diarrhoea (BVD), on the other hand, is a no less important agricultural disease~\cite{lanyon2013}, but does not have the same controversy surrounding control measures.

Our aim is to examine how language subjectivity on Twitter differs when comparing the discourses surrounding bTB and BVD. Using a combination of network analysis~\cite{newman2018networks} and language or sentiment analysis~\cite{liu2012}, we will examine the effect that the discourses have on the underlying network structures that connect conversations and users on Twitter, and how the types of user vary between them and who are the network ``influencers''. We hope that this will provide some insight into the characteristics of controversy and polarisation in disease-related discourse, and potentially provide avenues for surveillance and the combat of misinformation.

\section{Data and Methods}
The data used for this study was collected from the Twitter public API over the period from January 2018 to January 2021. Each set (bTB / BVD) has been collected by searching for suitable terms using the Twitter API search interface.

For bTB the simple search term ``bovine tb'' returned good coverage of the topic with apparently few off-topic tweets. However for BVD the acronym ``BVD'' returns too many other topics, whereas the full term ``bovine viral diarrhoea'' is rarely used. A compromise which returns much of the relevant discussion is the hashtag ``\#BVDFree''. BVDFree is the name of the name of the BVD control scheme in England~\cite{bvdfreeengland2022}. To build the BVD data set we first collected tweets with ``\#BVDFree'', then counted the most frequent words that appear alongside ``BVD''. The final search term includes ``bovine viral diarrhoea'', ``bvdfree'', ``stampoutbvd'', or ``BVD'' along with any of: ``herd'', ``virus'', ``cattle'', ``vet'', ``disease'', or ``farmer''.

The two data sets have been each used to compute three networks representing different aspects of Twitter discourse. In each aspect the vertices (nodes) represent Twitter users and the edges (links) represent: where one user retweets another, where one user replies to another, and where one user follows another; respectively we refer to these as the retweet, reply, and follow networks.

Using these networks and the contents of the tweets we identified the most influential users by background, nature, job, and disposition and applied text and sentiment analysis to classify the content and polarity (positive, negative, neutral) of each tweet. User influence was analysed using the common network measures of node in-degree and betweenness centrality~\cite{freeman1978}. Community clusters within the networks were computed using the Girvan-Newman algorithm~\cite{girvan2002}. The networks were visualised using the Gephi graph visualisation tool~\cite{associationgephi2022}.

We then investigated the content and language of each tweet. Word frequencies were counted to identify trends in topic and the vocabulary used across the two sets. Sentiment analysis was then performed over the two sets to determine the overall emotion and degree of polarisation present in the language used. Sentiment analysis was performed using the Vader sentiment analyser~\cite{hutto2014}. Vader is specifically designed for the language structure and vocabulary used in social media microblogging platforms and is extensively validated in this domain.

Finally we underwent a user tagging exercise where a number of volunteers categorised the high-influence users in each set. The top 100 users by each of in-degree and betweenness centrality were identified. These high-influence users were then tagged with a number of attributes: perceived occupation, whether they were a individual or an organisation, and whether the user's profile was largely focused on the single topic of either bTB or BVD. Each user was tagged by two volunteers, and overall agreement was checked and validated between taggers.

\section{Results}
\subsection{Network structure}
\begin{figure}
  \centering
  \small bTB tweet volume
  \includegraphics[width=\textwidth]{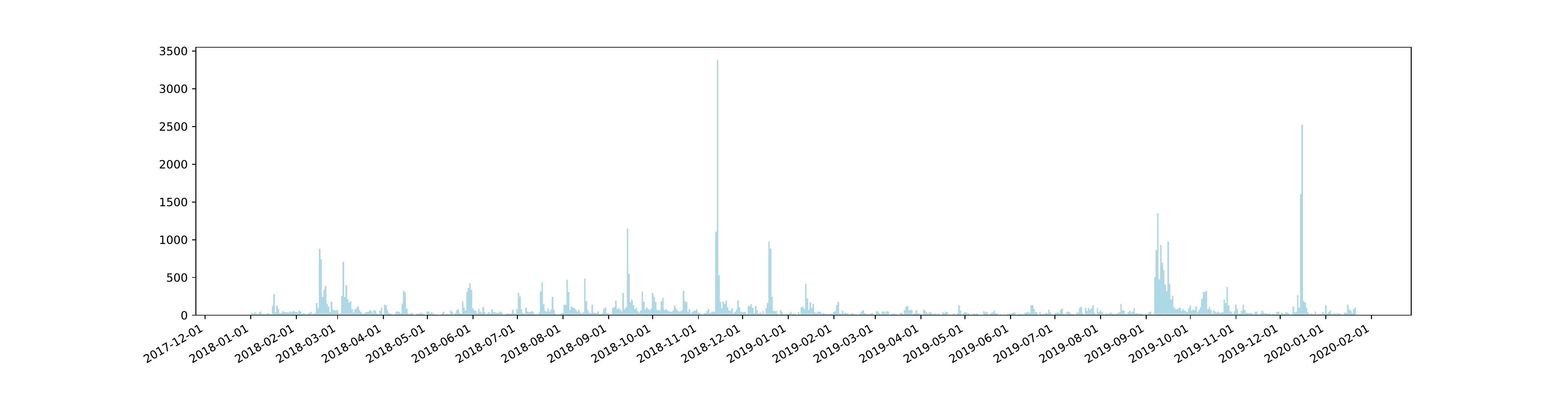}\\
  \small BVD tweet volume
  \includegraphics[width=\textwidth]{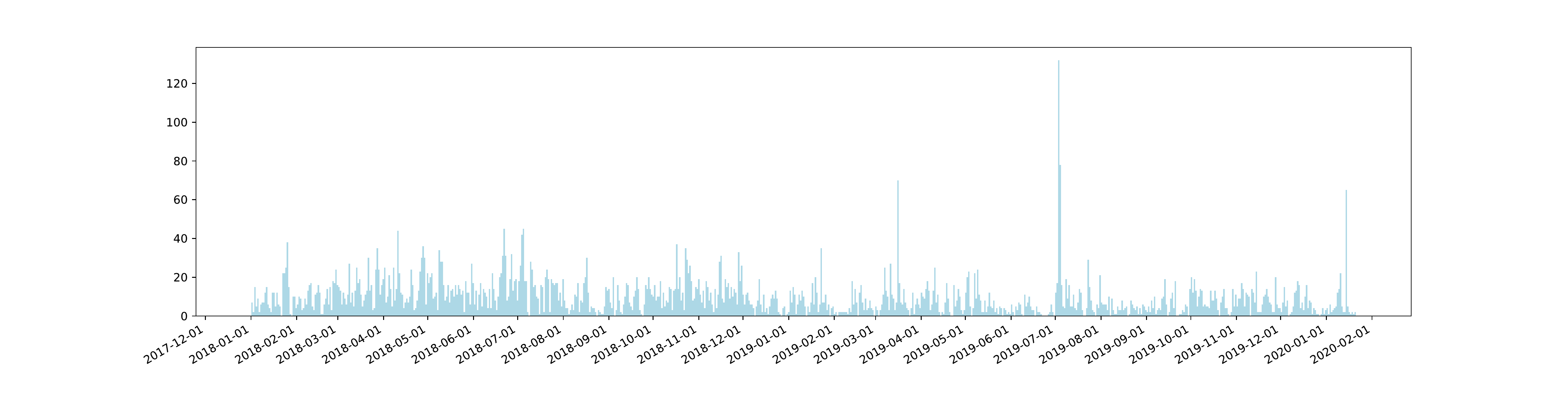}
  \caption{Volume of tweets over time for bTB (top) and BVD (bottom) between January 2018 and January 2020.}
  \label{fig:volume}
\end{figure}

Over the time period the search result for the bTB set contained 67,936 tweets and for the BVD set contained 7,290 tweets. Figure~\ref{fig:volume} shows the volume of tweets over time for each set.

The bTB set had higher volume overall, but with a number of very large volume spikes around significant news events For example, one such significant news event was around the release of the Godfray report, the UK government's Bovine TB strategy review, on 13th November 2018~\cite{godfray2018}. The BVD tweets have a much lower overall volume, but are also much more regular and less event driven.

Of the bTB tweets 80.8\% were retweets, 6.6\% were quoted tweets, and 4.0\% were replies to other tweets. In contrast, of the BVD tweets 63.8\% were retweets, 11.7\% were quoted tweets, and 7.4\% were replies. This suggests that despite the lower volume of tweets for BVD, there is more interaction or conversation around tweets. Whereas for bTB a much greater proportion are retweeted and amplified without more meaningful interaction.

\begin{figure}
  \centering
  %\hl{Fig removed for compile speed}
  \includegraphics[width=\textwidth]{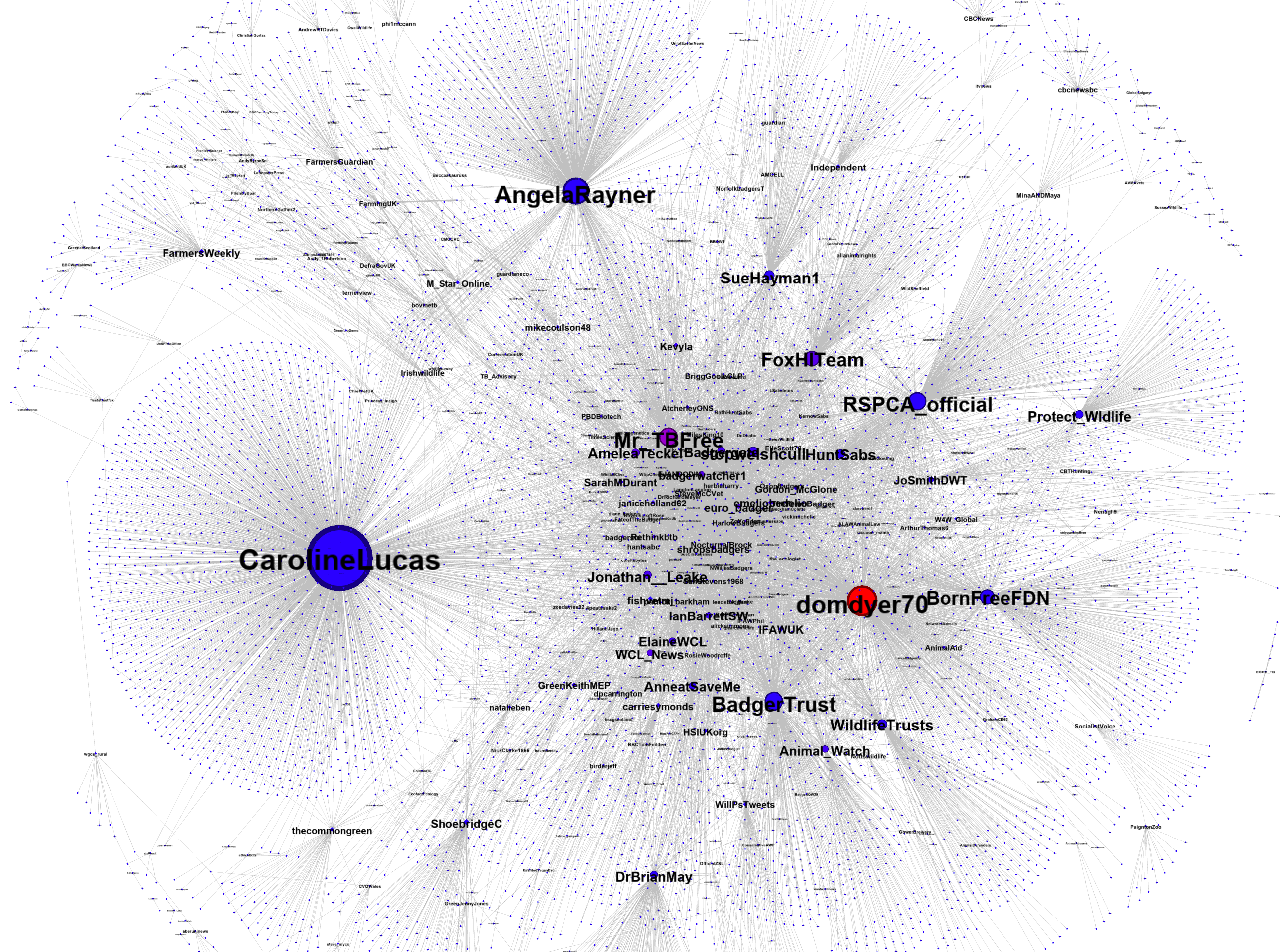}
  \caption{Visualisation of the bTB retweet network. Node size indicates the user's in-degree and colour indicates betweenness centrality (from low in blue, to high in red).}
  \label{fig:bTB-viz}
\end{figure}

\begin{figure}
  \centering
  %\hl{Fig removed for compile speed}
  \includegraphics[width=\textwidth]{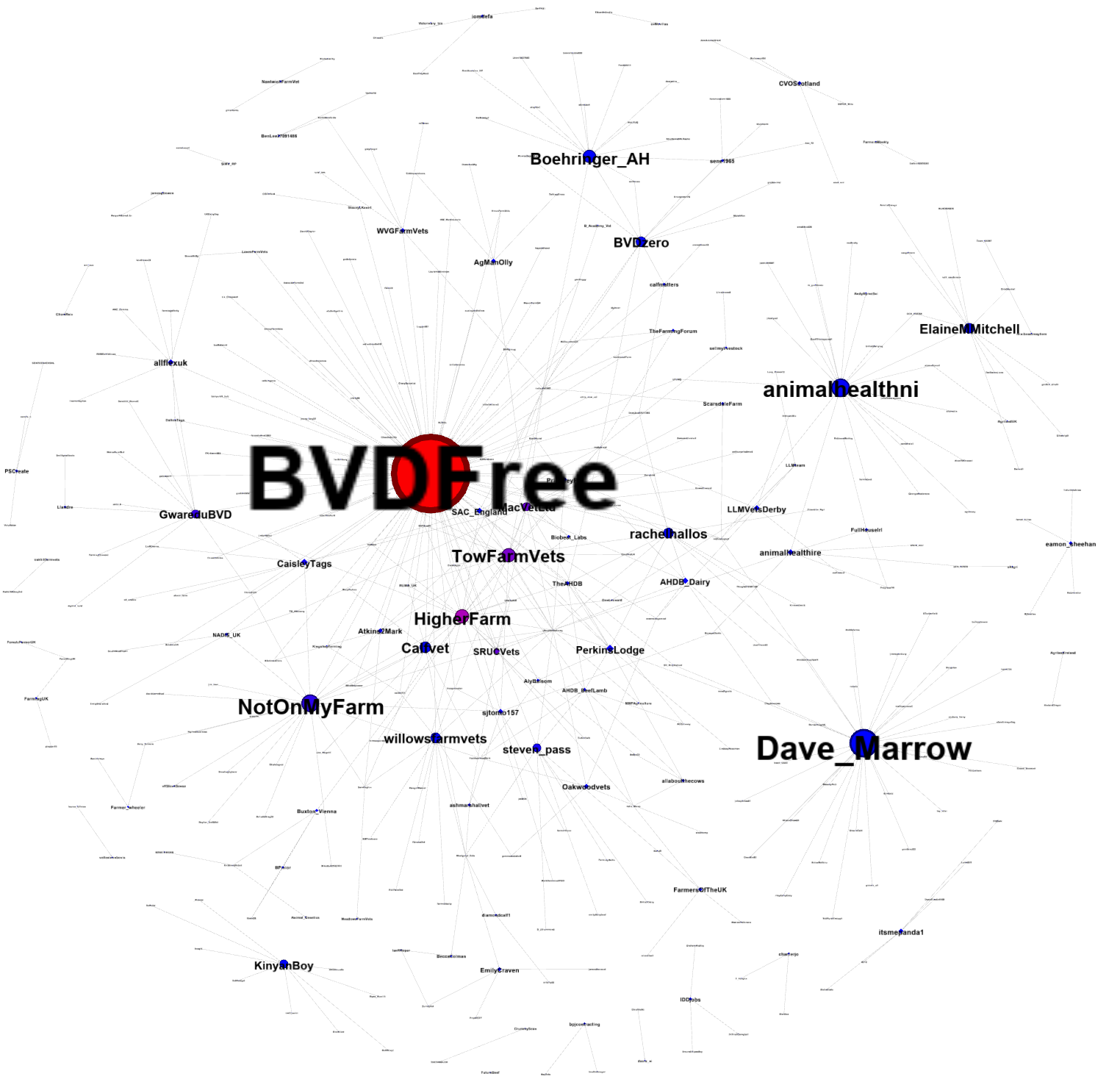}
  \caption{Visualisation of the BVD retweet network. Node size indicates the user's in-degree and colour indicates betweenness centrality (from low in blue, to high in red).}
  \label{fig:BVD-viz}
\end{figure}

The ranking of users in the networks by in-degree and by betweenness centrality identified two distinct types of influencer. Figures~\ref{fig:bTB-viz}~and~\ref{fig:BVD-viz} show visualisations of the retweet networks of each set; from this the two types of influencer can be observed. The first type of influencer is identified by high in-degree. The second type of influencer is identified by high betweenness centrality. The two networks show similar structure, despite the sparseness of the BVD network.

\begin{figure}
  \centering
  %\hl{Fig removed for compile speed}
  \includegraphics[width=\textwidth]{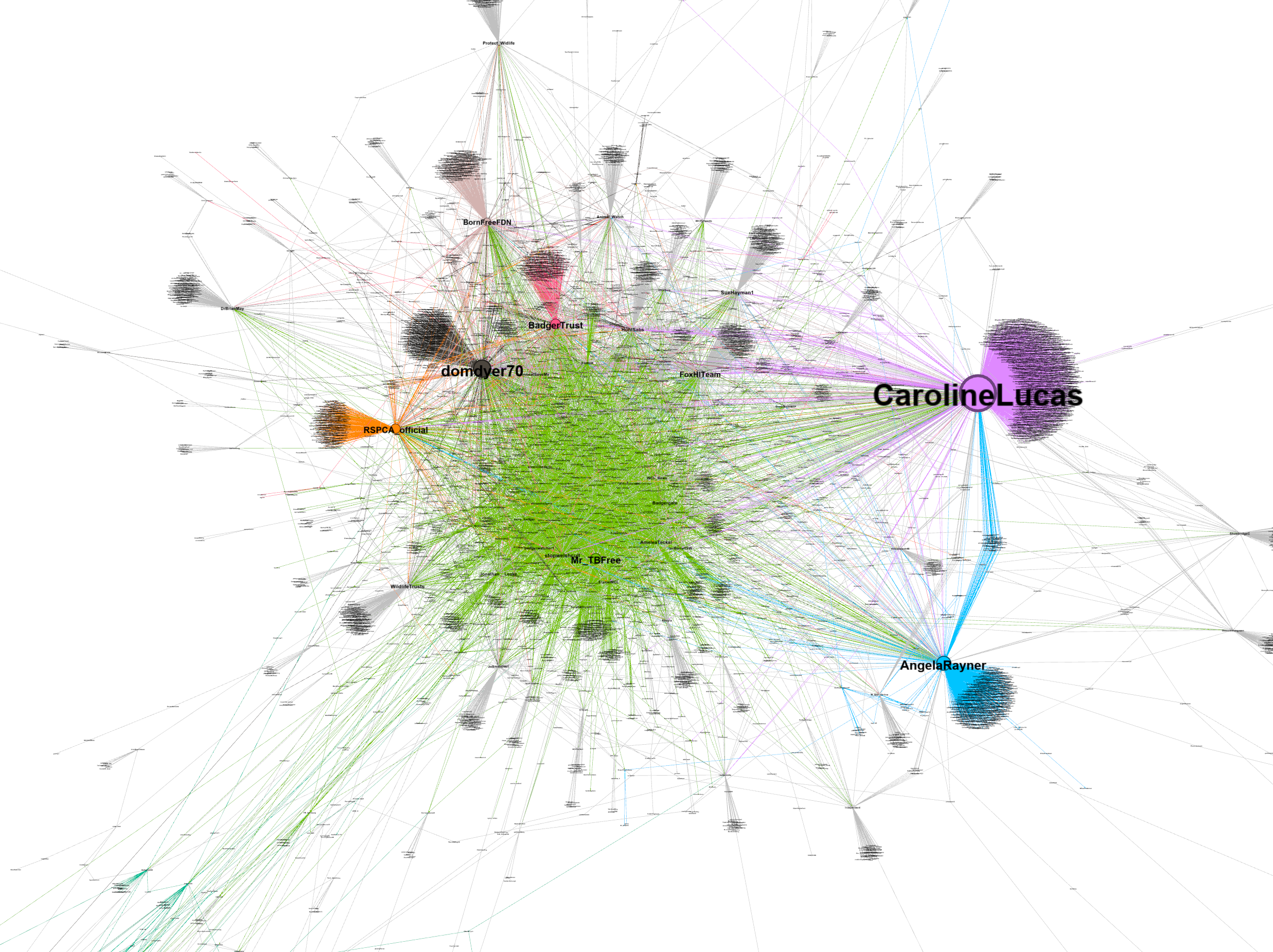}
  \caption{Girvan-Newman community clustering of the bTB retweet network. The large green cluser identifies the central community surrounding the topic, other colours identify sub-communities around the high-influence users.}
  \label{fig:bTB-cluster}
\end{figure}

\begin{figure}
  \centering
  %\hl{Fig removed for compile speed}
  \includegraphics[width=\textwidth]{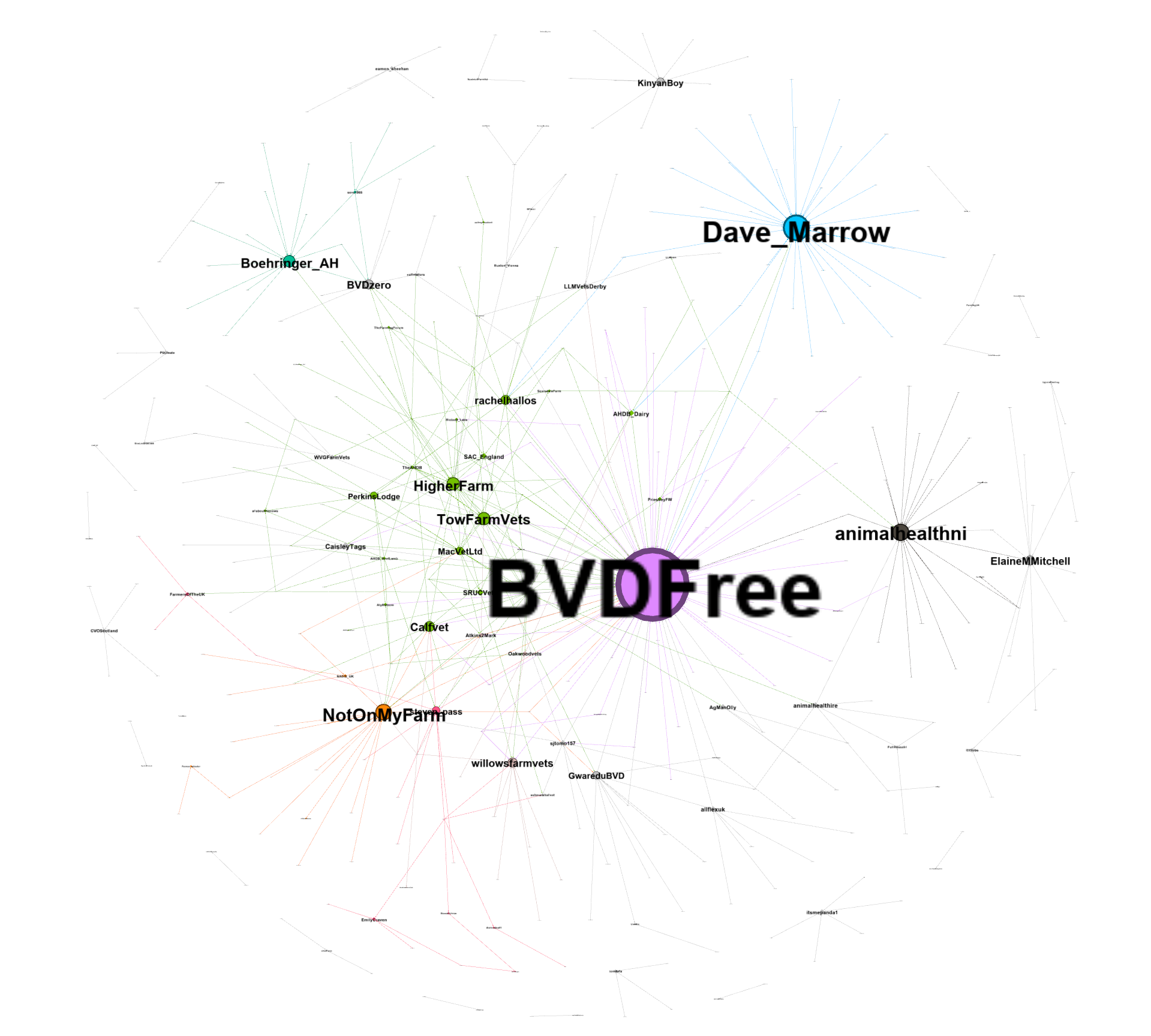}
  \caption{Girvan-Newman community clustering of the BVD retweet network. The large green cluser identifies the central community surrounding the topic, other colours identify sub-communities around the high-influence users.}
  \label{fig:BVD-cluster}
\end{figure}

The application of a community detection algorithm to the networks revealed the users forming the core community around each topic, and also a means to classify the sub-networks surrounding the high-influence users. Figures~\ref{fig:bTB-cluster}~and~\ref{fig:BVD-cluster} show visualisations of the community clustered networks.

A comparison of distributions of degree and centrality measures for each network revealed no notable difference between the two topics.

\subsection{Language and sentiment}
\begin{figure}
  \centering
  \includegraphics[width=0.49\textwidth]{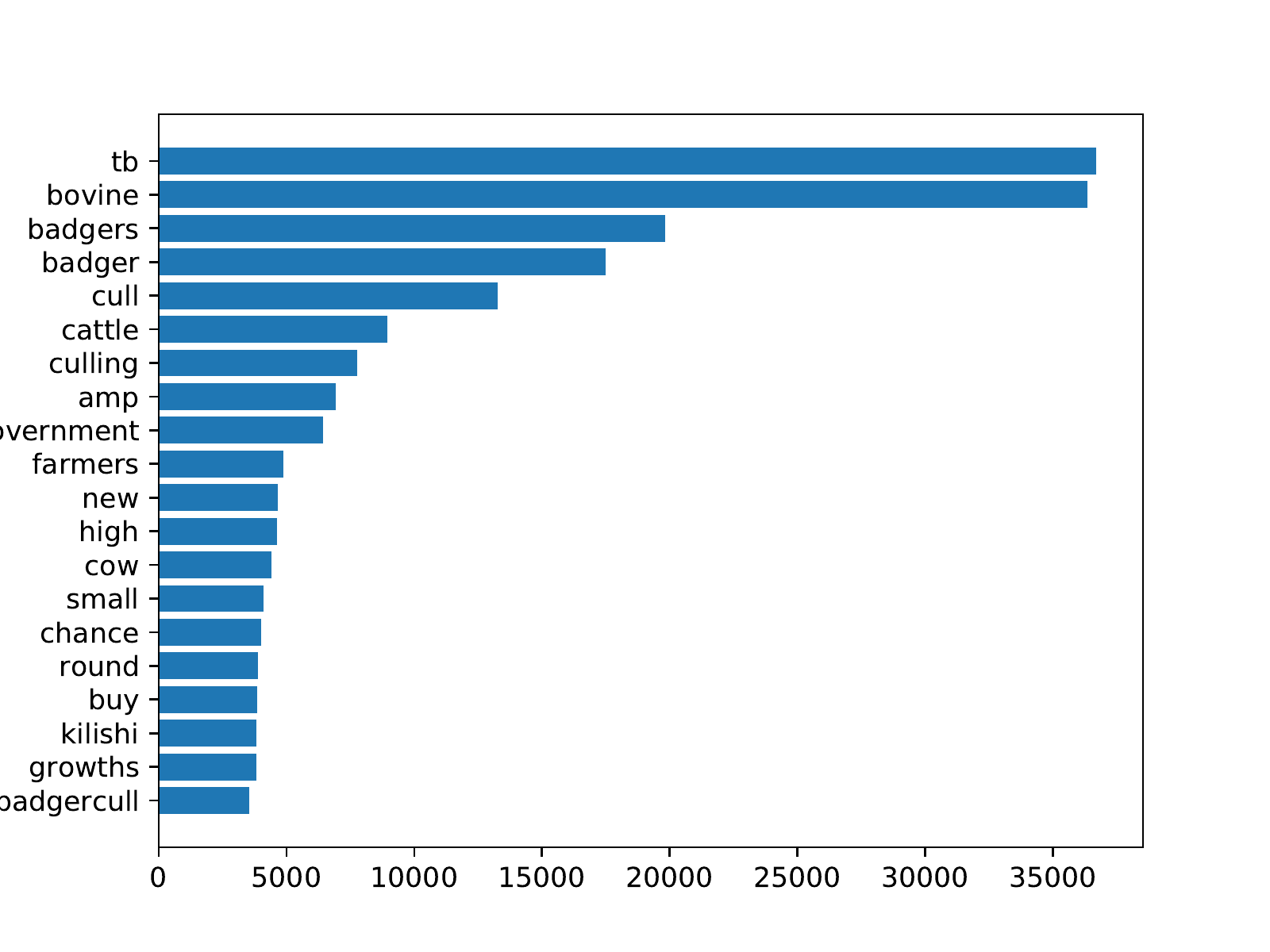}
  \includegraphics[width=0.49\textwidth]{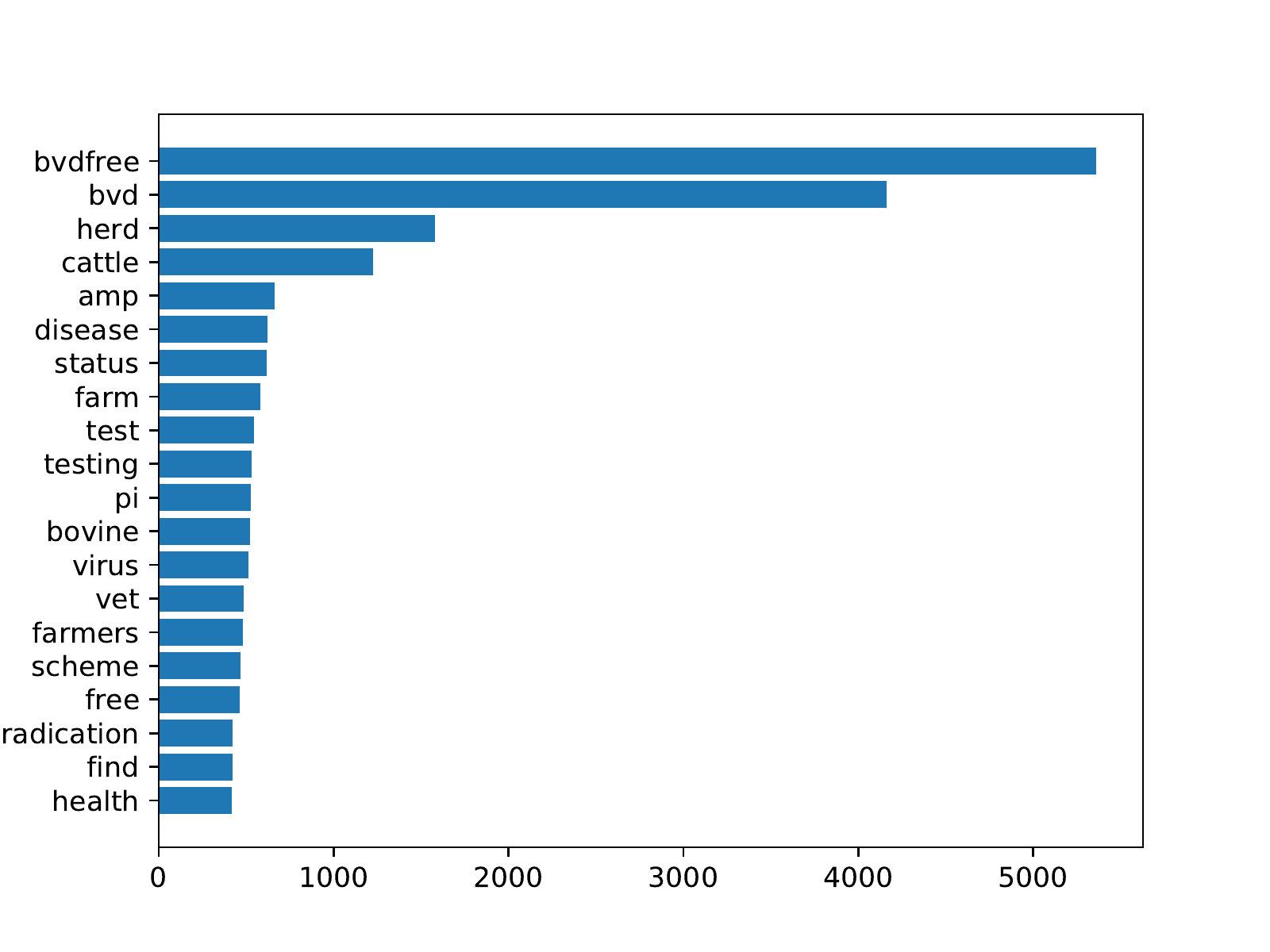}
  \caption{Word frequencies contained within bTB tweets (left) and BVD tweets (right).}
  \label{fig:wordfreq}
\end{figure}

We then examined the content of the tweets themselves. First we counted word frequencies within each topic. Figure~\ref{fig:wordfreq} shows the most common terms within each set. For BVD the most common terms were either directly disease related or agricultural in nature: e.g.\ herd, cattle, disease, status, farm. For bTB, however, the most frequent terms are dominated by those relating to badgers and culling: e.g.\ badgers, cull, culling, government.

\begin{figure}
  \centering
  \includegraphics[width=0.49\textwidth]{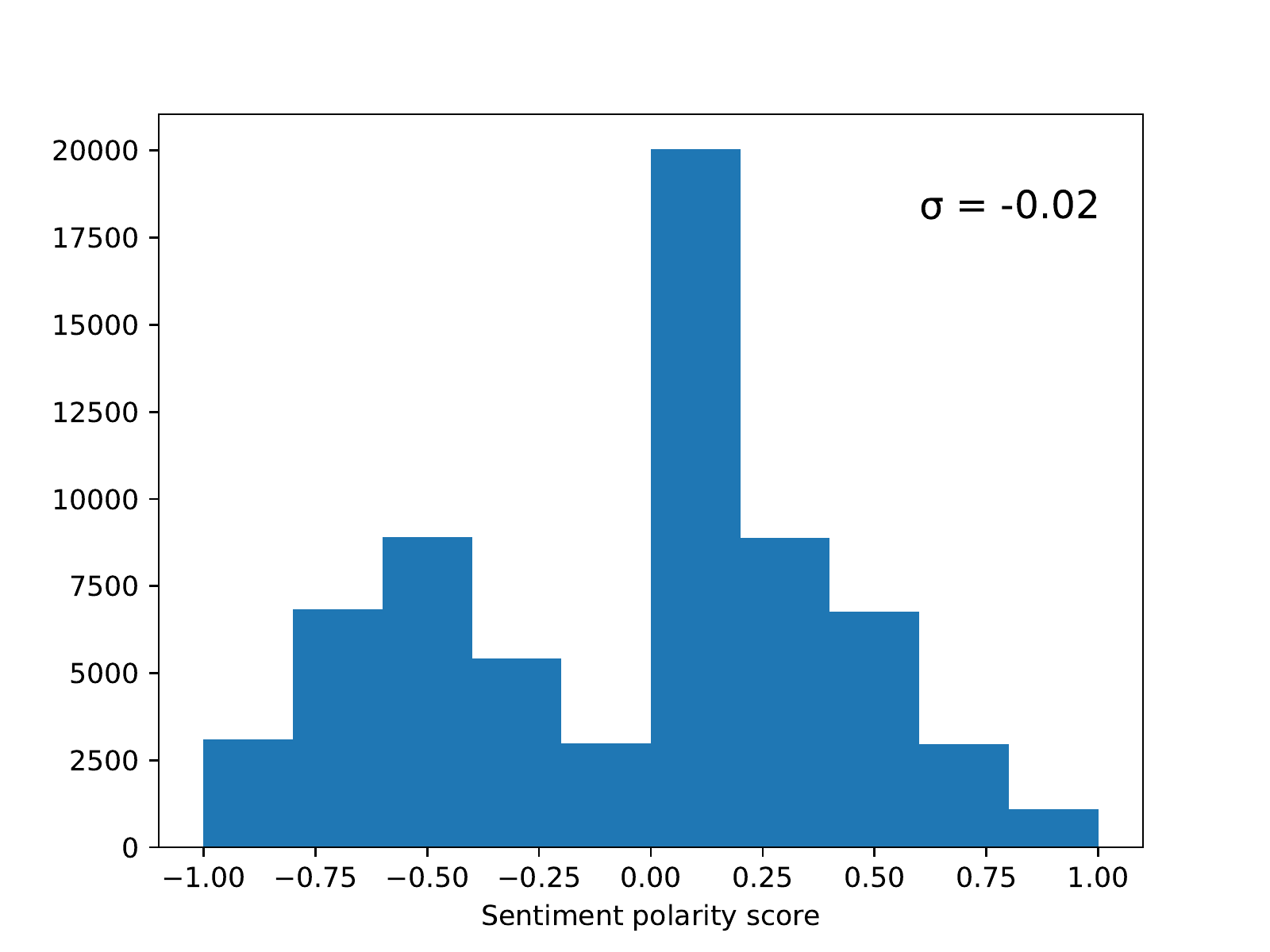}
  \includegraphics[width=0.49\textwidth]{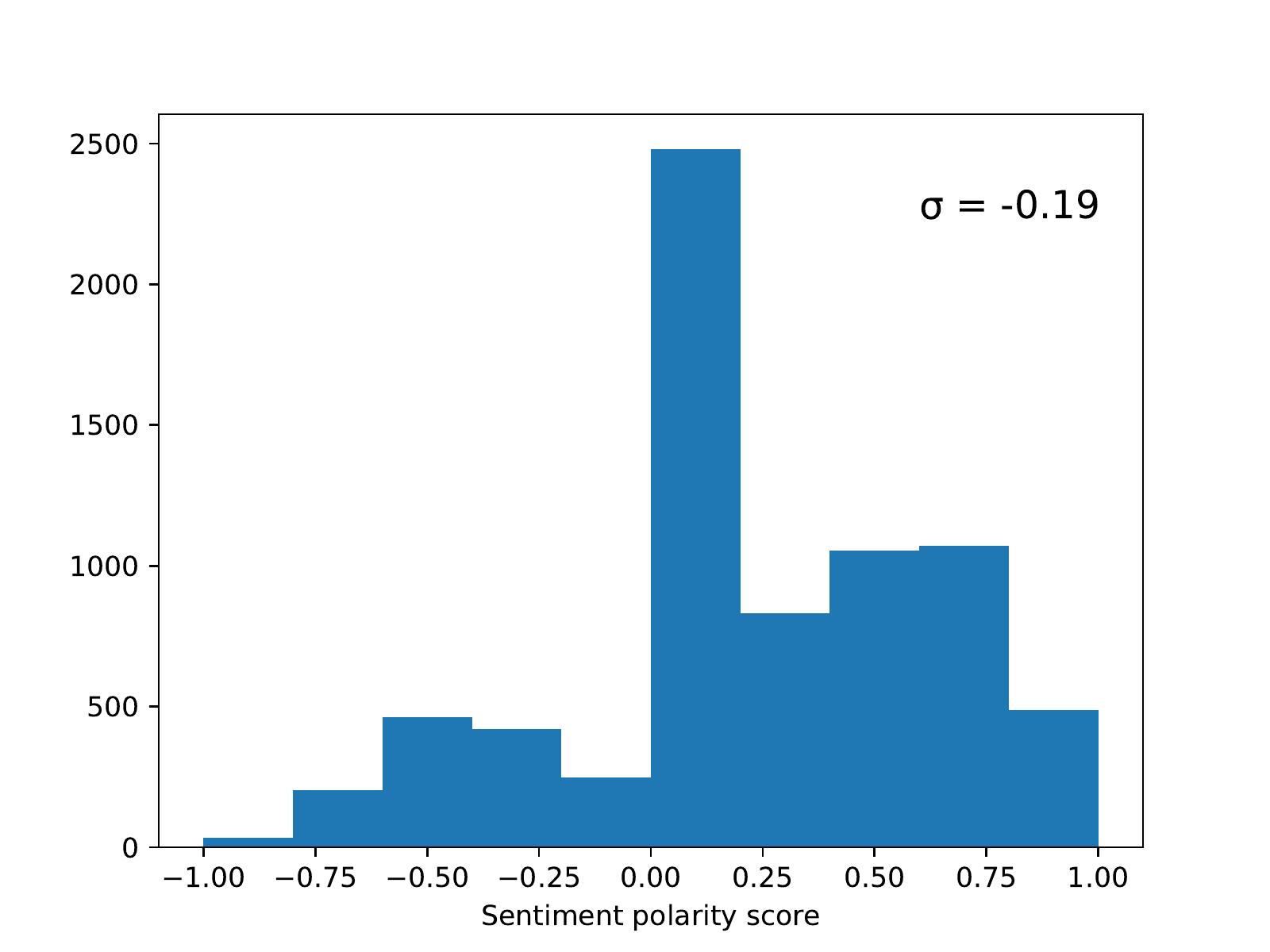}
  \caption{Sentiment score distribution for bTB tweets (left) and BVD tweets (right). A more negative skew $\sigma$ indicates more positive sentiment in general.}
  \label{fig:sentiment}
\end{figure}

Performing sentiment analysis on each set revealed further differences. The sentiment analysis algorithm computed a sentiment score for each tweet, with a positive score representing more positive sentiment and a negative score representing more negative sentiment. The magnitude of the score represents the level of emotion identified. Figure~\ref{fig:sentiment} shows the sentiment score distributions for each set. Both sets on the whole show largely neutral content (0 score), however the bTB set shows a greater proportion of tweets with negative sentiment than BVD. BVD tweet sentiment is skewed more heavily to the positive, indicating much more generally positive sentiment. 

\begin{figure}
  \centering
  \small bTB tweet sentiment\\
  \includegraphics[width=\textwidth]{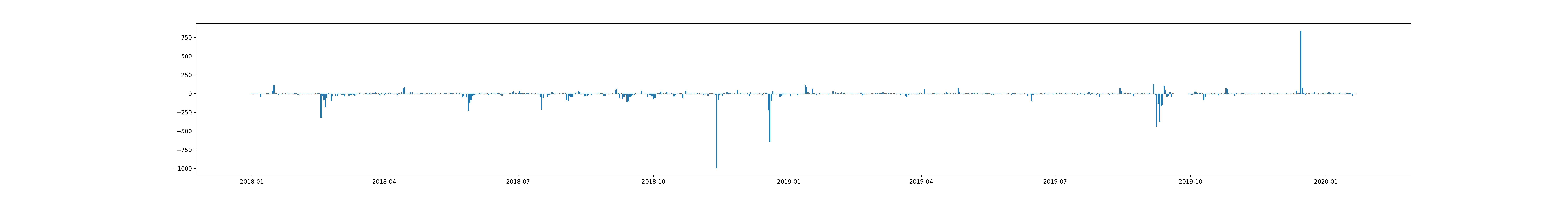}\\
  \small BVD tweet sentiment\\
  \includegraphics[width=\textwidth]{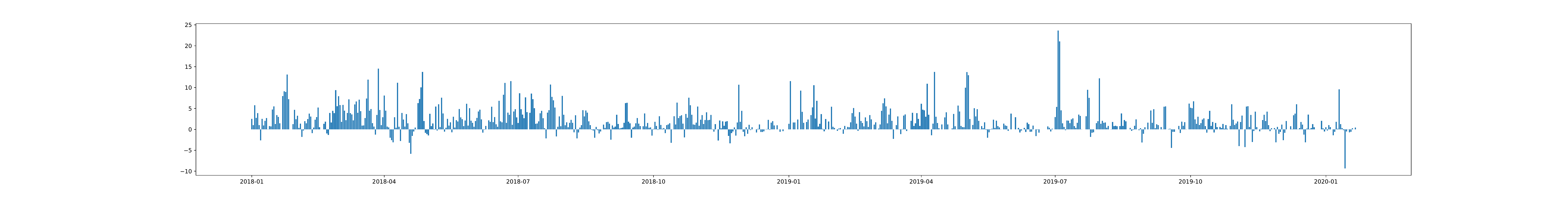}
  \caption{Total daily sentiment score time series for bTB tweets (top) and BVD tweets (bottom).}
  \label{fig:sentiment-time}
\end{figure}

Analysing sentiment over time revealed more about the event driven nature of sentiment. Figure~\ref{fig:sentiment-time} shows the total daily sentiment score for each set. The BVD set shows much more daily positive tweeting, with only a few spikes in sentiment. However, the bTB set shows more fluctuation between positive an negative, with large spikes in sentiment on some days. The large sentiments spikes correspond with news events, like the previously identified release of the Godfray report.

\begin{figure}
  \centering
  %\hl{Fig removed for compile speed}
  \includegraphics[width=0.49\textwidth]{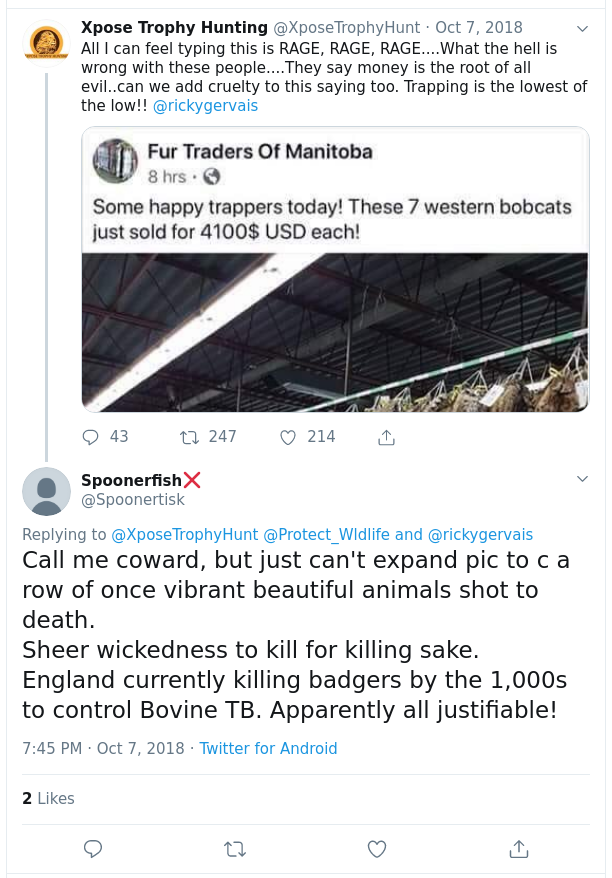}
  \includegraphics[width=0.49\textwidth]{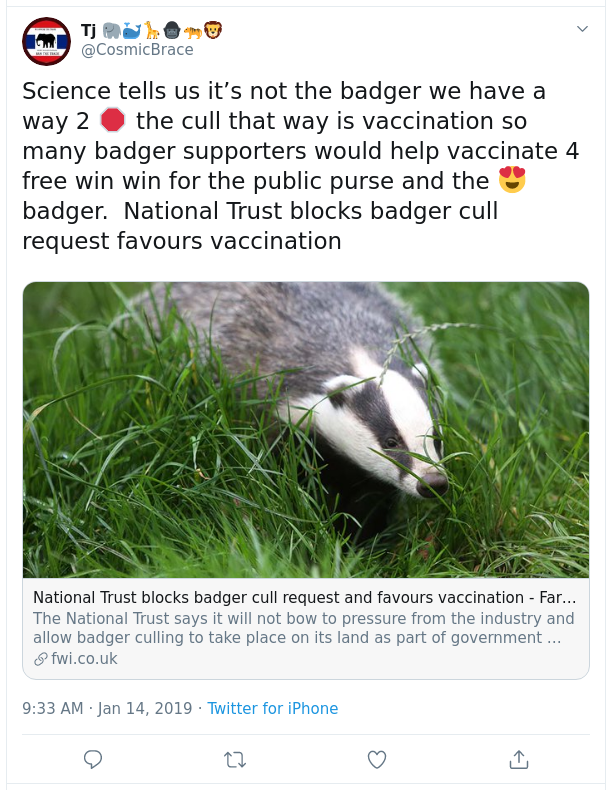}\\
  \rule{\textwidth}{0.2pt}
  \includegraphics[width=0.49\textwidth]{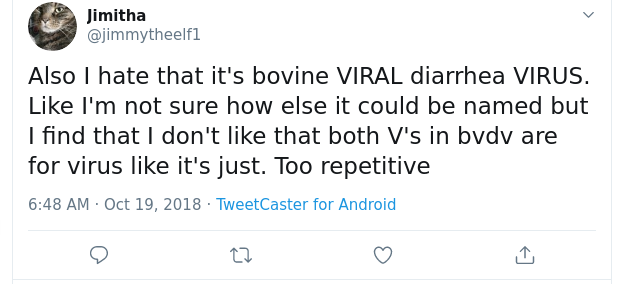}
  \includegraphics[width=0.49\textwidth]{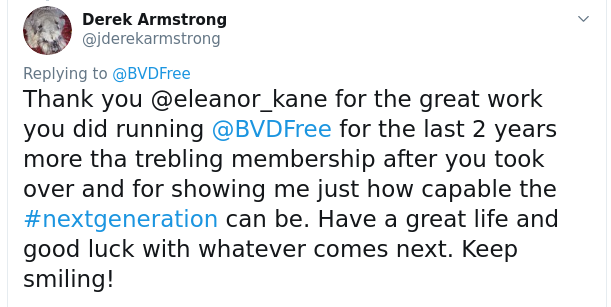}
  \caption{Examples of tweets that score highly for either negative (left) or positive (right) sentiment, on bTB (top) and BVD (bottom).}
  \label{fig:sentiment-tweets}
\end{figure}

Examples of tweets that score highly for either negative or positive sentiment, in each set, are shown in Figure~\ref{fig:sentiment-tweets}.

\subsection{User characteristics}
The results of the user tagging exercise reveal differences in the types of actor engaged in each topic. Table~\ref{tab:user-tags} shows the breakdown of identified user types within the top 100 users identified by either in-degree or betweenness centrality. BVD has spread of mainly vets, farmers, and academics tweeting about it, whereas bTB tweeters are dominated by activists and conservationists. Politicians are much more engaged with bTB than BVD. Those who tweet about bTB tend to be individuals, whereas BVD has a mix of individuals and organisations. Of those organisations tweeting about BVD, many are commercial entities, whereas relatively few commercial entities are tweeting about bTB. 22\% of the users tweeting about bTB are solely focused on the topic of bTB, but 96\% of BVD tweeters have other interests.

Each tag applied to a user was decided by at least two volunteers independently, and there was an overall agreement between volunteers of 93.5\%.

\begin{table}
  \centering
  \begin{tabular}{|l|r||l|r|}
    \hline
    bTB & \% & BVD & \% \\
    \hline\hline
    Activist &                  42.9 & Vet                      & 39.2\\
    Conservationist &          37.9 & Farmer                   & 26.4\\
    Academic                 & 13.7 & Academic                 & 17.6\\
    Politician            &     7.7 & News Source              &  7.2 \\
    News Source               & 7.1& Politician               &  1.6 \\
    Vet                       & 6.6&  Activist                 &  1.6 \\
    Farmer                 &    6.0  &  Conservationist        &    0.0   \\
    \hline
    Individual            &    62.1&  Organisation / group    &   52.8\\
    Organisation / group  &    34.1&  Individual              &   45.6 \\
    \hline
    Commercial?          &     0.5 & Commercial?             & 37.5 \\
    \hline
    bTB-centric?& 22 & BVD-centric? & 4 \\
    \hline
  \end{tabular}
  \caption{Breakdown of user types in each network, identified by volunteers in a user tagging exercise.}
  \label{tab:user-tags}
\end{table}

\section{Conclusions}
While analysing network structure showed little difference between the two disease topics, elements of the structure allowed us to better investigate the more language and user-centred differences between the two topics. Using network measures we identified two types of influencer with distinct roles, identified by different measures. Users with high in-degree are retweeted by a large number of other users, meaning their reach within the global Twitter network is amplified and their tweets are more widely seen in general. Users with high betweenness centrality are more central to the specific disease network (as opposed to more generally in the global Twitter network), they form bridges between other users within the disease network. These users can therefore be seen to be the influencers who are most embedded with in the topic. Users can, of course, be both types of influencer and those with both types will be the most influential within their specific disease network.

Sentiment was the best indicator of polarisation and controversy, with bTB tweets showing greater proportions of negative and very negative sentiment. The skew in the distribution of sentiment for BVD was a good indicator of the lack of controversial material. It was also clear that outpourings of more extreme sentiment, both positive and negative, were driven by occasional events rather than a constant flow.

Using our knowledge of the different types of influencer in the network, we were able to identify the major types of actor involved in each network. This revealed a number of differences between the two diseases and the users who form the network surrounding the topic. Vets, farmers, and academics, are the most engaged with the topic of BVD, communicating largely in neutral language, even with a more positive outlook. bTB discourse is heavily dominated by wildlife conservationists and activists who often use more extremely negative emotive language---a clear indicator of the controversy surrounding the subject of badger culling.

This analysis only scratches the surface of characterising indicators of polarisation and controversy in these disease-related topics and there is plenty of further work in this area. One could imagine, for example, that machine learning and natural language processing techniques could aid in the user characterisation process, or that the role of bots in the discourse should be investigated. However, we have identified some of the essential characteristics that will enable a better understanding of polarisation and controversy in disease discourse, and potentially open avenues for surveillance or the combat of misinformation.

%%%%%%%%%%%%%%%%%%%%%%% 
%%%%% References %%%%%%
%%%%%%%%%%%%%%%%%%%%%%% 

\bibliographystyle{unsrt}
\bibliography{twitter}

%%%%%%%%%%%%%%%%%%%%%%%
%%%%% License    %%%%%%
%%%%%%%%%%%%%%%%%%%%%%%
\vspace{1.5em}
\noindent For the purpose of open access, the author has applied a Creative Commons Attribution (CC~BY) licence to any Author Accepted Manuscript version arising from this submission.

\end{document}